\documentclass[sigconf]{acmart}

\AtBeginDocument{%
  \providecommand\BibTeX{{%
    \normalfont B\kern-0.5em{\scshape i\kern-0.25em b}\kern-0.8em\TeX}}}

\setcopyright{acmcopyright}
\copyrightyear{2020}
\acmYear{2020}

\acmConference[SIGGRAPH '20]{SIGGRAPH '20}{July 19--23, 2020}{Washington, DC}
\acmBooktitle{SIGGRAPH '20,
  July 19--23, 2020, Washington, DC}
\acmPrice{15.00}
\acmISBN{978-1-4503-XXXX-X/18/06}

\begin{document}

\title{MAGES 3.0: Tying the knot of medical VR}


\author{George Papagiannakis}
\affiliation{
\institution{University of Crete, ICS - FORTH,}
\institution{ORamaVR}}
\email{george.papagiannakis@oramavr.com}

\author{Paul Zikas}
\affiliation{\institution{ORamaVR}}
\email{paul@oramavr.com}

\author{Nick Lydatakis}
\affiliation{\institution{ORamaVR}}
\email{nick@oramavr.com}

\author{Steve Kateros}
\affiliation{\institution{ORamaVR}}
\email{steve@oramavr.com}

\author{Mike Kentros}
\affiliation{\institution{ORamaVR}}
\email{mike@oramavr.com}

\author{Efstratios Geronikolakis}
\affiliation{\institution{ORamaVR}}
\email{stratos@oramavr.com}

\author{Manos Kamarianakis}
\affiliation{\institution{ORamaVR}}
\email{manos.kamarianakis@oramavr.com}

\author{Ioanna Kartsonaki}
\affiliation{\institution{ORamaVR}}
\email{ioanna.kartsonaki@oramavr.com}

\author{Giannis Evangelou}
\affiliation{\institution{ORamaVR}}
\email{giannis.evangelou@oramavr.com}

\renewcommand{\shortauthors}{George Papagiannakis, et al.}

\begin{abstract}
In this work, we present MAGES 3.0, a novel Virtual Reality (VR)-based authoring SDK platform for accelerated surgical training and assessment. The MAGES Software Development Kit (SDK) allows code-free prototyping of any VR psychomotor simulation of medical operations by medical professionals, who urgently need a tool to solve the issue of outdated medical training. Our platform encapsulates the following novel algorithmic techniques: a) collaborative networking layer with Geometric Algebra (GA) interpolation engine b) supervised machine learning analytics module for real-time recommendations and user profiling c) GA deformable cutting and tearing algorithm d) on-the-go configurable soft body simulation for deformable surfaces.\end{abstract}

\maketitle

\section{Introduction}
There is a growing lack of medical professionals globally and not enough already are trained today for future needs \cite{Papagiannakis2018}. The number of new surgeons trained per year has not changed in the last 30 years, whereas the population has doubled. There is an urgent need for a paradigm shift in medical training to overcome the challenges.

Previously we have proven that MAGES \cite{MAGES} makes medical training more efficient. In a revolutionary clinical study \cite{HOOPER2019} in cooperation with New York University that established - for the first time in the medical bibliography - skill transfer and skill generalisation from VR to the real Operating Room in a quantifiable, measurable ROI.

Utilizing the new advances in MAGES 3.0, we released four new medical VR training modules: a dental implant placement, an endotracheal intubation, a series of emergency medical scenarios and a REBOA operation.

\section{The MAGES 3.0 technology}
In the following sections we showcase our MAGES 3.0 innovations.

\subsection{Multiplayer with GA interpolation (M)}
Our networking layer proposes a low bandwidth and high visual fidelity collaborative module, featuring multiple users based on our proprietary virtual character interpolation engine. We deploy a unique representation of the 3D deformable meshes in our Geometric Algebra (GA) framework that allows 4x improvement on reduced data network transfer and lower CPU/GPU usage for this task. This allows for a high number of multiple concurrent users in the same collaborative virtual environment.

Participants can join the same virtual OR to communicate and interact while completing the training scenario. Even non-medical related users are able to learn and perform basic surgical steps while assisted from our personalized recommendation system. 

\subsection{Analytics based on ML agent with recommendations (A)}
In medical training it is crucial to provide a user assessment capable of reflecting the educational impact of the used methodology. We designed an analytics system to fulfil this need.
Our analytics engine features a Machine Learning (ML) agent, facilitating a virtual surgeon supervising the training session, collecting surgical errors, user decisions and time intervals. We train our ML agent from medical experts, constructing a unique trainee profile to propose real-time suggestions to users according to their level of experience. Each surgical action is segmented into individual parts to identify interactions such as tool handling and proper usage of medical equipment. Our supervised ML model is capable to understand the validity of each action and decide whether to offer assistance with additional audio-visual guidance.

\begin{figure}
  \includegraphics[width=\linewidth]{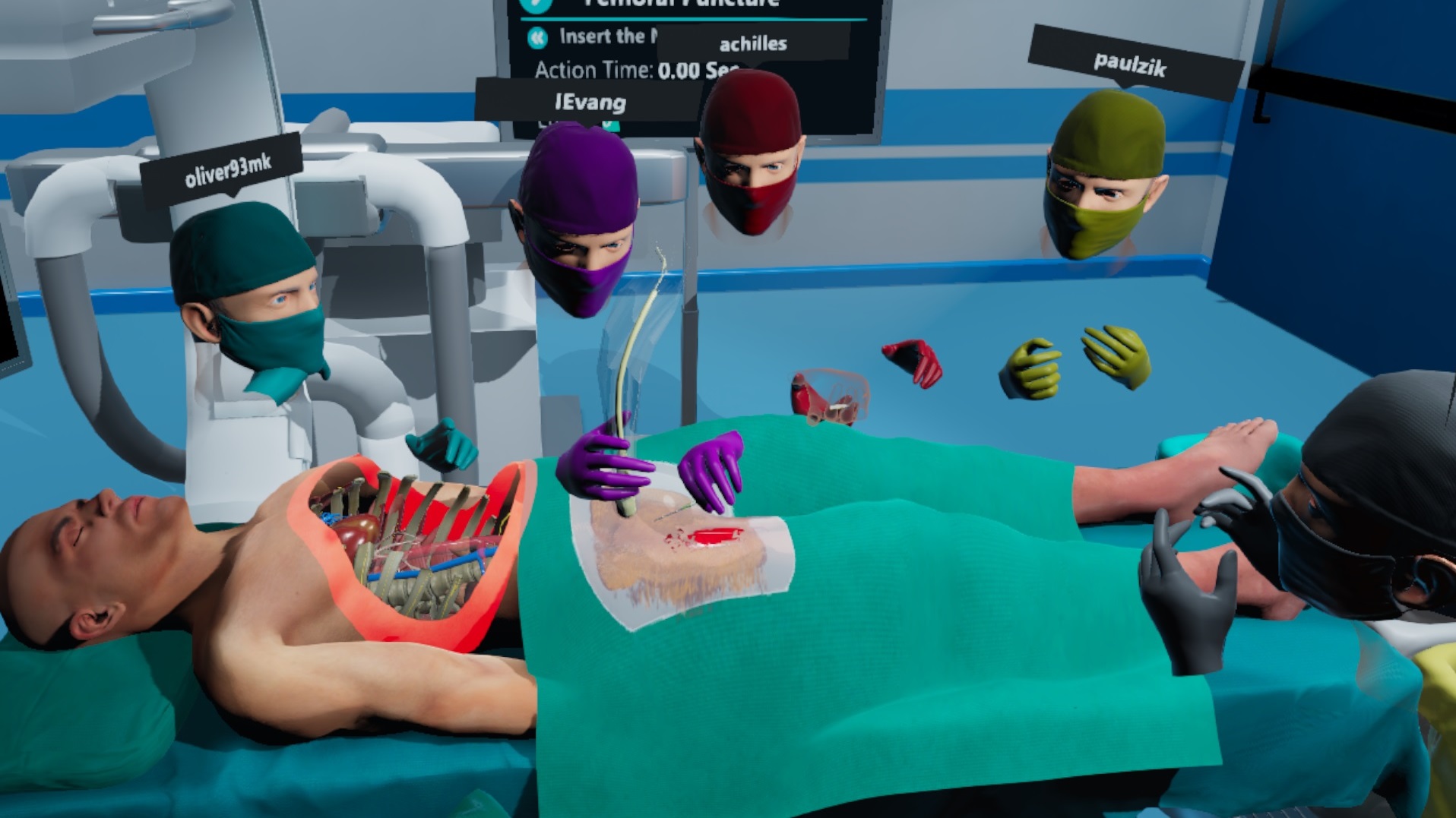}
  \caption{A cooperative REBOA training scenario in easy mode where users can visualize the patient's arteries.}
  \label{fig:teaser}
\end{figure}

\subsection{Geometric Algebra deformable animation, cutting and tearing (G)}
Our work focuses in enhancing the state-of-the-art in skinning and handling of models. The use of quaternions and dual quaternions yielded fast results, free of interpolation problems or other geometric artifacts. Another step towards that direction is the introduction of Conformal Geometric Algebra (CGA). In a CGA framework, all points (P), translations (T), rotations (R) and dilations (D) are described as a single entity, called multivectors. Since the interpolation of two multivectors of type $A\in\{P,T,R,D\}$ yields a multivector of the same type, we acquire an easier to understand and implement interpolation algorithm without the need to constantly transmute objects of two worlds such as in the case of quaternion-matrices.

The CGA framework we implemented converts the original skinning equation to a multivector-only equivalent:
\begin{equation}\label{eq:cga_formula}
C_k[m] = \displaystyle
\sum_{n\in I_m} w_{m,n}(M_{n,k}B_{n})c[m](M_{n,k}B_{n})^\star
\end{equation}

\noindent where $B_n$ amounts to an offset matrix, $c[m]$ is the  image of the $m$-th vertex of the model in CGA, $w_{m,n}$ is the influence of the $n$-th bone on the $m$-th vertex, $M_{n,k}$ is the transformation of the $n$-th bone at time $k$, $I_m$ is the list of bones influencing the $m$-th vertex and $C_k[m]$ is the image of the $m$-th vertex of the final model at time $k$.

The latter implementation yields animations close to the former method, while also enabling us to perform character deformable cutting and tearing. Furthermore, our engine performs animations with less intermediate keyframes, reducing bandwidth.

\subsection{Editor in VR (E)}
Our SDK encapsulates a VR editor capable of generating VR training scenarios following our modular Rapid Prototyping architecture. We designed our system as a collection of authoring tools combining a visual scripting system and an embedded VR editor forming a bridge from product conceptualization to product realization and development in a reasonably fast manner without the fuss of complex programming and fixtures. 
MAGES SDK platform is designed for any programmer or doctor to make the development of various surgical scenarios rapid and simple. Therefore the platform allows for non-VR experts to develop new surgical modules/scenarios or modify existing ones, increasing the platform's possibilities.

\subsection{Semantically annotated soft bodies (S)}
In the core of MAGES lies an advanced mathematical algorithm for physics-based visual techniques to allow the 3D representation of deformable soft body objects (skin, tissue, etc.), essential for VR surgical training. Since surgical training is all about cutting and suturing soft body objects, collision detection (touching) and handling of soft bodies with other objects becomes crucial for high-realism VR.
The soft deformation algorithm is based on shape matching techniques and particle-based soft body simulations. Our methodology differs from the state of the art since it provides on-the-go control of the particles as physical objects and a centre point, which controls the entire soft body. Velocity based interaction can be applied directly to the corresponding particles while interacting with the environment as objects. In addition we synchronize the deformable objects over network utilizing our GA interpolation engine to improve interaction with concurrent users.

Finally, we extended our physics engine to support interaction with ropes and even giving user the ability to perform knots and sutures in VR. In combination with our soft body mechanics we developed a bowel anastomosis operation where users can interact with the virtual sutures and soft tissue.  

\section{Conclusions and Future Work}
In this work, we presented the main innovations of MAGES 3.0 featuring a VR authoring solution for medical training. By improving medical training and access to high-quality medical care, MAGES apps have the potential to improve treatment outcomes globally. Through efficient and affordable training of medical professionals, we can significantly increase the number of people with access to medical care, saving lives, costs, time.

We are currently working with edge computing resources to cover more geographies and users, thus truly enabling collaborative training that will benefit from 5G advancements. 

\begin{acks}
This project has received funding from the European Union's Horizon 2020 research and innovation programme under grant agreement No 871793 (ACCORDION) and No727585 (STARS-PCP) and supported by Greek national funds (projects VRADA and vipGPU).
\end{acks}

\bibliographystyle{ACM-Reference-Format}
\bibliography{sample-sigconf}

\end{document}